\documentclass[iop]{emulateapj}

\usepackage[colorlinks=true,citecolor=blue,linkcolor=blue]{hyperref}

\shorttitle{Dragonfly Nearby Galaxies Survey. I.}
\shortauthors{Merritt et al.}

\begin{document}

\title{The Dragonfly Nearby Galaxies Survey. I. Substantial variation in the diffuse stellar halos around spiral galaxies}

\author{Allison Merritt\altaffilmark{1}, Pieter van
  Dokkum\altaffilmark{1}, Roberto Abraham\altaffilmark{2,3} \& Jielai Zhang\altaffilmark{2,3}}

\altaffiltext{1}{Department of Astronomy, Yale University, New Haven, CT, USA}
\altaffiltext{2}{Department of Astronomy and Astrophysics, University of 
   Toronto, Toronto, ON, Canada}
\altaffiltext{3}{Dunlap Institute, University of Toronto}

\begin{abstract}
Galaxies are thought to grow through accretion; as less massive galaxies are disrupted and merge over time, their debris results in diffuse, clumpy stellar halos enveloping the central galaxy. Here we present a study of the variation in the stellar halos of galaxies, using data from the Dragonfly Nearby Galaxies Survey (DNGS). The survey consists of wide field, deep ($\mu_{g} > 31$ mag arcsec$^{-2}$) optical imaging of nearby galaxies using the Dragonfly Telephoto Array. Our sample includes eight spiral galaxies with stellar masses similar to that of the Milky Way, inclinations of $16-90$ degrees and distances between $7-18$ Mpc. We construct stellar mass surface density profiles from the observed $g$-band surface brightness in combination with the $g-r$ color as a function of radius, and compute the halo fractions from the excess stellar mass (relative to a disk$+$bulge fit) beyond $5$ half-mass radii. We find a mean halo fraction of $0.009 \pm 0.005$ and a large RMS scatter of $1.01^{+0.9}_{-0.26}$ dex. The peak-to-peak scatter is a factor of $>100$ -- while some galaxies feature strongly structured halos resembling that of M31, three of the eight have halos that are completely undetected in our data. We conclude that spiral galaxies as a class exhibit a rich variety in stellar halo properties, implying that their assembly histories have been highly non-uniform. We find no convincing evidence for an environmental or stellar mass dependence of the halo fraction in the sample. 
\end{abstract}


\section{Introduction}
In a $\Lambda$CDM universe, the hierarchical growth of dark matter
structures is mirrored by the buildup of galaxies through a series of
mergers and low mass accretion events
\citep{Purcell2007,Johnston2008,deLucia2008,Cooper2013,Pillepich2014}. Stars 
stripped from satellite galaxies during this process come to reside
in diffuse and highly structured stellar halos enveloping the central galaxy,
which, as a consequence of the relatively long dynamical timescales in the
outskirts of galaxies, retain a ``memory'' of past accretion events. 
Understanding the accretion histories of individual galaxies, then, amounts to
characterizing their stellar halos.

The recent non-detection of the stellar halo of the massive spiral
galaxy M101 down to $32$ mag arcsec$^{-2}$ by \cite{vanDokkum2014}
stands in contrast to the extensively studied stellar halos of the
Milky Way
\citep[e.g.,][]{Majewski2003,Belokurov2006,McConnachie2006,Carollo2007,Bell2008}
and M31
\citep[e.g.,][]{Ibata2001,Ferguson2002,Ibata2007,Richardson2008,McConnachie2009,Gilbert2012},
and strongly suggests that the variation between the stellar halos of
massive spiral galaxies may be greater than previously thought.  

Despite results from several studies which demonstrate that the
scatter in the stellar populations
\citep{Seth2005,Mouhcine2007,Monachesi2013,Monachesi2015}, amount of
substructure
\citep{BlandHawthorn2005,Mouhcine2010,MartinezDelgado2010,Miskolczi2011,RadburnSmith2012,Okamoto2015},
and even luminosity fractions
\citep{Shang1998,Barker2009,Tanaka2011,Vlajic2011,Barker2012,Bakos2012}
of stellar halos are qualitatively consistent with expectations from
theory, the \textit{stellar masses} of these halos, which in principle
provide a robust point of comparison, in general remain
unexplored \citep[with only a few exceptions outside of the Local Group,
e.g. ][]{Seth2007,Bailin2011,Greggio2014,vanDokkum2014,Streich2015}. 

The main difficulty in measuring stellar masses is the need to characterize both
the global structure of the stellar halo and its constituent stellar populations.
Star counts surveys, which are capable of constructing color and metallicity profiles
for the stellar halos of nearby galaxies \citep[e.g.][]{Monachesi2015}, are typically
restricted by sparse area coverage at larger distances, and the resulting profiles
are thus susceptible to the presence of underlying substructure and not guaranteed to
reliably represent the halo on large scales. Integrated light surveys are more efficient,
but these observations are commonly plagued by scattered light from bright stars or the
dense centers of galaxies. Light thrown into the wings of the point spread function (PSF)
can masquerade as a stellar halo \citep{deJong2008,Sandin2014}, and must be carefully
controlled for or removed \citep{Slater2009,Duc2015}. Accordingly, measurements of 
stellar halo masses must rely on assumptions about at least one of these two components.
The few measurements of stellar halo masses that have been made are $-$ thus far $-$ also not
straightforward to compare, due to inconsistent methodology between different studies. 

A more comprehensive understanding of the scatter in stellar halo
mass fractions can be achieved through consistent measurements of the
stellar halos of a larger sample of spiral galaxies. In this paper we
present early results from the Dragonfly Nearby Galaxies Survey (DNGS). This is
an ongoing photometric survey of nearby galaxies carried out with the Dragonfly
Telephoto Array, a robotic refracting telescope optimized to cleanly detect
low surface brightness emission \cite{Abraham2014}. With its large field of
view ($\sim 2^{\circ} \times 3^{\circ}$), Dragonfly provides panoramic images of
each targeted galaxy.  

Using deep ($>30$ mag arcsec$^{-2}$) surface brightness and $g-r$
color profiles, we measure the stellar mass surface density profiles
for eight objectively selected spiral galaxies. We derive the stellar
halo masses for our sample from the surface density profiles and show
that, even within a narrow mass range, the stellar halo mass fractions
of spiral galaxies can vary by several orders of magnitude. 

\begin{figure*}[!t]
\begin{center}
\includegraphics[width=\textwidth]{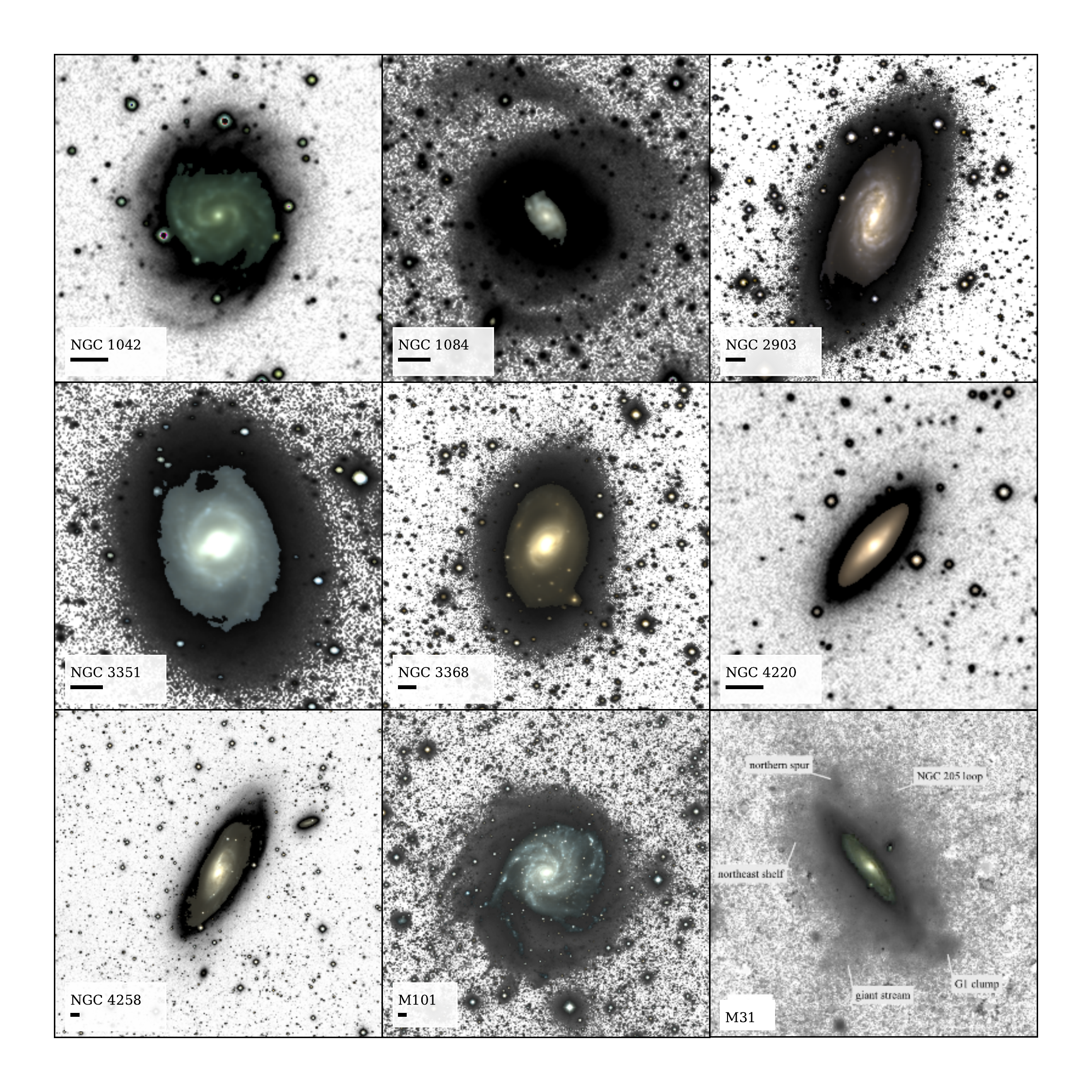}
\caption{Images of each of the eight galaxies in this sample. The
    pseudo-color images were created from $g$ and $r$ band images for
    the high surface brightness regions, and the greyscale shows the
    lower surface brightness outskirts. The bottom right panel shows
    M31, created from a combination of Dragonfly and PAndAS
    data \citep{McConnachie2009,Carlberg2011} and redshifted to a
    distance of 7 Mpc \citep{vanDokkum2014}. Black lines beneath each galaxy
    name indicate scales of 1 arcmin.  \label{allspirals}}
\end{center}
\end{figure*}

\vspace{2em}

\section{Data}
\subsection{Sample selection and observations}
We used the Dragonfly Telephoto
Array \citep{Abraham2014} to observe a sample of eight spiral
galaxies. The array was used in an 8-lens
configuration, prior to its upgrade to the present $48$-lens configuration.
The sample covers six separate Dragonfly fields, corresponding to a total
area of $\sim 55$ square degrees on the sky. The galaxies were selected on the
basis of their absolute magnitude ($M_{B} < -19.3$), and lie at distances of
$7-18$ Mpc. 

These fields were observed as part of an ongoing
photometric survey of nearby bright galaxies (DNGS). The selection criteria
for DNGS are based on absolute magnitude and distance, such that we
obtain volume-limited samples in different bins of absolute
magnitude. Effectively, in each absolute magnitude bin, we select the
nearest galaxies that are visible from the New Mexico Skies
Observatory. The only additional criterion is that the galaxies have
low Galactic cirrus in their immediate vicinity ($F_{100\mu\rm m} <
1.5$ mJy/Sr along the line of sight, from IRAS maps). We emphasize
that we did not target galaxies on the basis of known tidal debris or
any other physical property except absolute $B$ band luminosity. For
galaxy properties, see Table \ref{properties}. 

We observed our sample throughout 2013 and 2014. Typical per-night
exposure times were 5400\,s or 10,800\,s, taken in 9-point dither
patterns of  $600$s individual exposures. Offsets between dither
positions were either $25'$ or $40'$. Typical total exposure times
ranged from 15 to 20 hours per galaxy.

Considering the (very low) surface brightnesses of the features we intend to study, observing under optimal conditions is critical. To avoid the effects of scattered moonlight and significant atmospheric cirrus, we took care to observe only during clear, dark conditions. Frames with obvious cirrus-induced effects (manifesting as round ``halos'' surrounding bright stars) were removed from the final coadded images. Diffuse atmospheric cirrus is more difficult to avoid, in particular because it can drift unnoticed in and out of our field of view over the course of the night; however, we are able to identify its signature in our data by performing photometric calibrations on each individual frame \citep[using SDSS stars as reference][]{Abazajian2009} and excluding frames with discrepant photometric zeropoints on a given night. This procedure, which is remarkably effective, is described in detail in J. Zhang et al (in preparation).

\begin{figure*}[!t]
\begin{center}
\includegraphics[width=\textwidth]{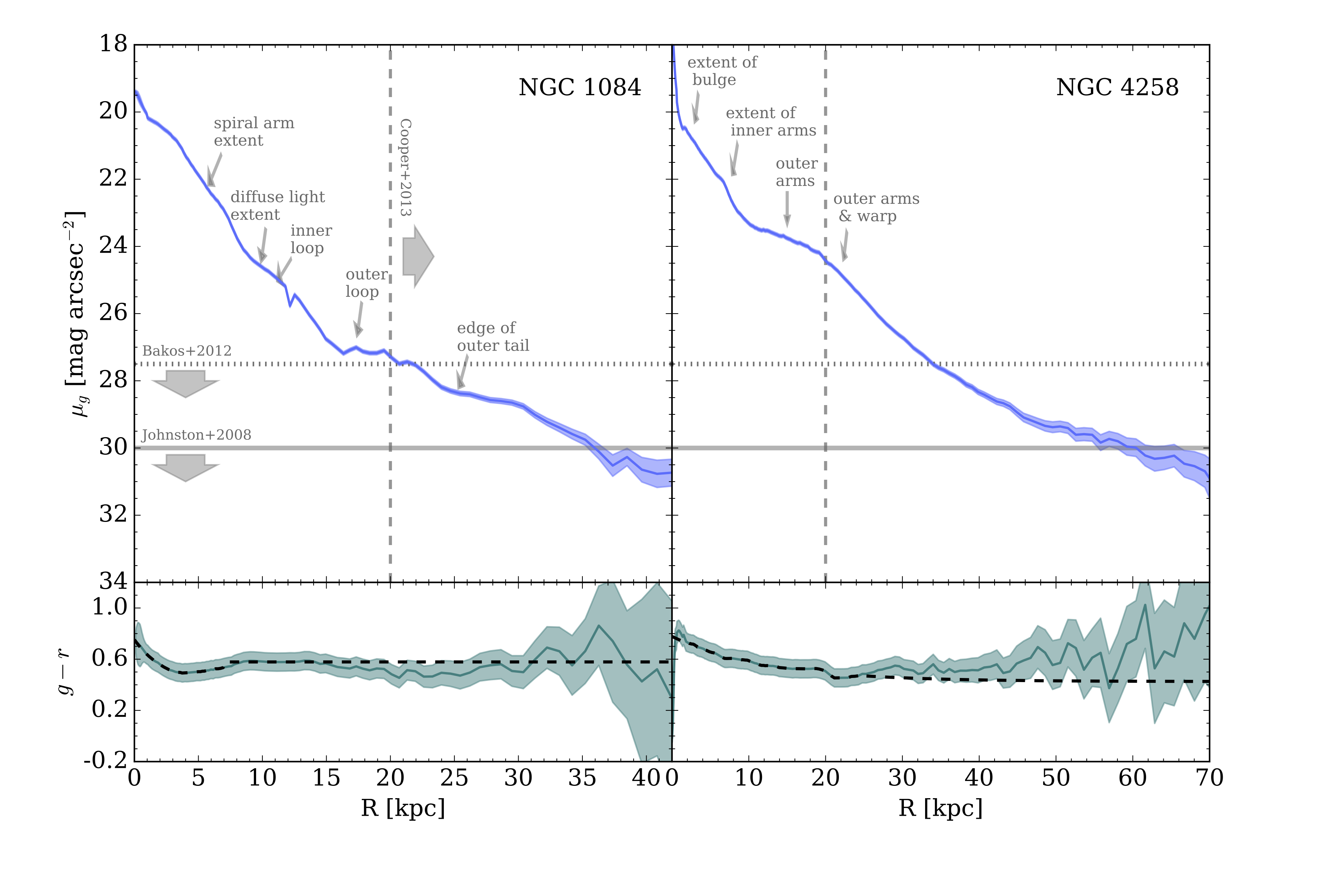}
\caption{
    \textit{Top panels}: Two example surface brightness profiles (and
    $1 \sigma$ error bars) with
    labeled morphological features as a function of distance along the
    semi-major axis. Solid horizontal lines highlight the
    requisite depth to detect faint streams \citep{Johnston2008}, and
    dotted horizontal lines indicate the surface brightness limit
    below which M31-like halos are expected to
    dominate \citep{Bakos2012}. The dashed vertical grey line shows the
    radius beyond which stellar halos are expected to dominate the
    light profile \citep{Cooper2013}. The features in the profile of NGC
    1084  beyond $\sim 7$ kpc correspond to the stellar halo, whereas
    the profile of NGC 4258 is driven by the spiral disk out beyond
    $\sim 20$ kpc. \textit{Bottom panels}:  Observed and best fit $g-r$ color
    profiles. The shaded region around the observed color profiles
    represents the $1\sigma$ error. \label{egmug}}
\end{center}
\end{figure*}

\begin{figure*}[!t]
\begin{center}
\includegraphics[width=\textwidth]{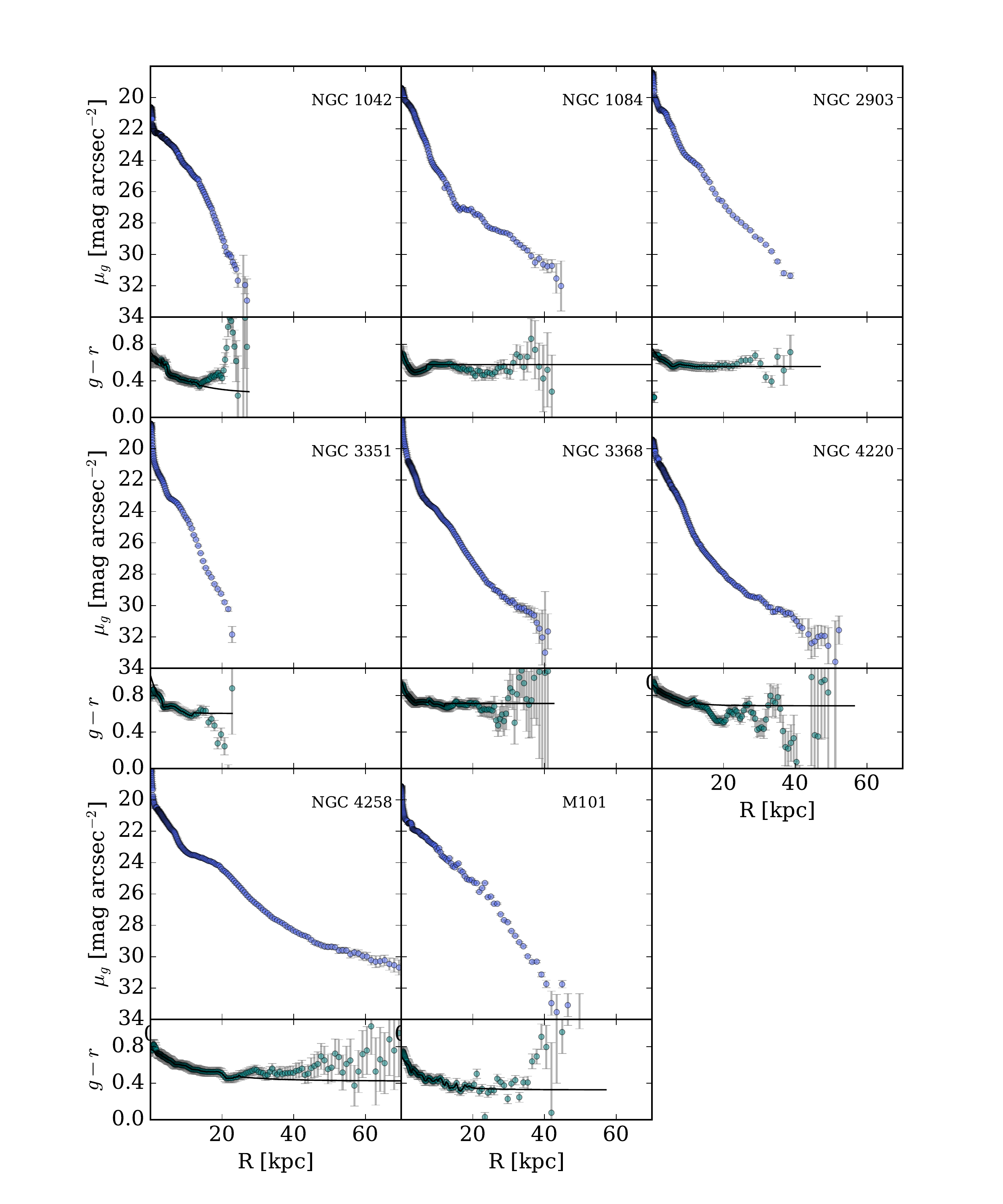}
\caption{Same as Figure \ref{egmug}, but for the entire
    sample.  \label{mug}} 
\end{center}
\end{figure*}

\subsection{Background variations and PSF effects}
After dark subtraction and flat fielding, the data show residual gradients
in the background of each individual frame,
which are largely due to the changing sky background with zenith
distance. We modeled these gradients in each individual frame with a
second order polynomial, and subtracted them. All objects in the field were
aggressively masked
to prevent their flux from affecting the fit. Owing to the large
dithers and the stability of the instrument, no further steps were
necessary to flatten the images \citep[further details can be
found in][]{vanDokkum2014}. 

Star subtraction was performed using a custom python pipeline that
computes an empirical, spatially-varying composite PSF from Dragonfly
images. First, we divided the image into $n \times n$ subframes
($3\times3$ proved to be sufficient). An average unsaturated PSF was
constructed in each subframe from a large number of unsaturated
stars. To ensure that no galaxies or overly elongated stars were 
included in the PSF, we iteratively rejected individual candidate
sources that deviated signficantly from the average PSF. Furthermore,
flux from all neighboring objects was masked out during the
average. For each star, the central region of the composite PSF was
thus constructed from the unsaturated PSFs. To ensure smooth variation over
the field, at any ($x,y$) position in the frame, the model PSF was a weighted
average of the PSFs in the nearest subframes. The wings of the PSF were built from
all saturated stars in the field (i.e., we did not account for spatial
variations in the wings). 

After background subtraction and star subtraction, the $1\sigma$
surface brightness limits of each final, coadded $g-$band frame are
$29-29.8$ mag arcsec$^{-2}$ in $60" \times 60$" boxes; or
$28.6-29.2$ mag arcsec$^{-2}$ in $12"$ regions (the size of a $6\times
6$ binned pixel). To arrive at these values, we placed 100 boxes (of the aforementioned size)
down in empty regions of our data and measured the variation in the average flux contained
therein. In this study, however, we integrated the flux in $6\times 6$ binned images
over broad elliptical annuli,
and as the sample galaxies span up to $\sim 10$ arcmin, we reach magnitudes
of $31-32$ mag arcsec$^{-2}$ in the surface brightness profiles. The actual limit
is not the same in each field, as it depends on the number of foreground stars, the size
of the galaxy, and the amount of foreground Galactic cirrus. The derivation of
the errorbars in our profiles is discussed in Section 3.

\section{Results}

The Dragonfly images of the eight spiral galaxies are shown in Figure \ref{allspirals}.
Each image is $30$ arcmin on a side. The familiar high surface brightness regions of
the galaxies are shown in color; the low surface brightness outskirts are shown
in grey scale. The galaxies show a remarkable variation in their low surface brightness
outskirts $-$ NGC 1084 has a nearly spherical halo featuring a giant tidal tail, whereas
both NGC 1042 and M101 show spiral arms down to very faint surface brightness levels.
In the bottom right, we show what M31 looks like when it is shifted out to a distance
of $7$ Mpc (comparable to M101) and placed in an empty region of a Dragonfly field
\citep[see][]{vanDokkum2014}. M31-like halos, characterized by significant substructure,
are easily detectable in our data but appear to be atypical. In the following subsection,
we quantify this observed variation in the stellar halos of Milky Way mass spiral
galaxies using the surface brightness profiles of the galaxies.

\subsection{Surface brightness and stellar mass surface density
  profiles}

\begin{figure*}[!t]
\begin{center}
\includegraphics[width=\textwidth]{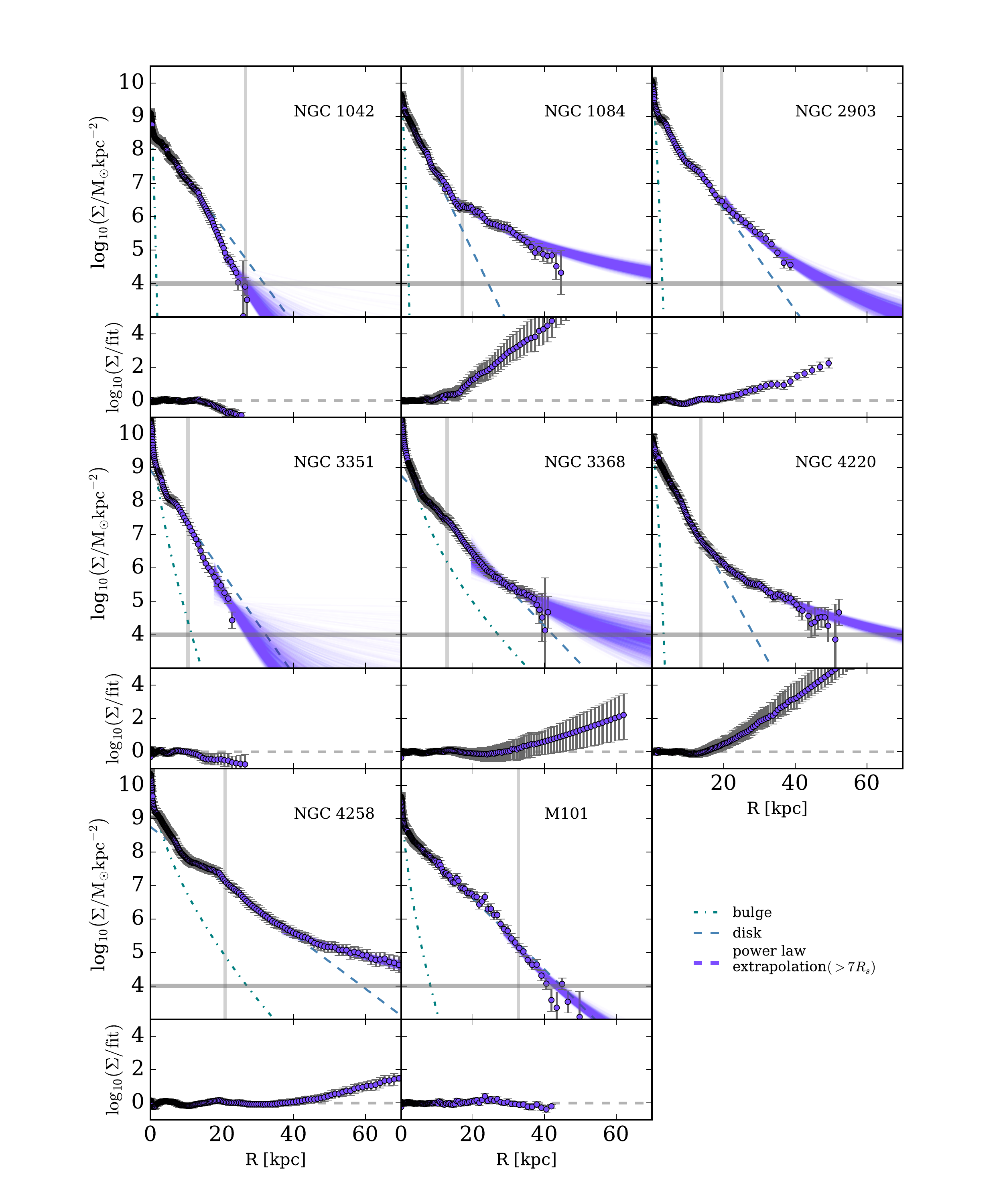}
\caption{
    \textit{Top panels}: Purple data points show the measured surface density
    profiles (and $1 \sigma$ error bars). Dashed and faded solid
    purple lines indicate an extrapolation of this profile beyond
    $7R_{s}$ (vertical line) down to $10^{4}M_{\odot}$ kpc $^{-2}$
    (horizontal line). Best fit models of the disk and bulge are shown
    in green dot-dashed and blue dashed lines, respectively. See text
    for details. \textit{Bottom panels}: Residuals after subtracting the bulge 
    and disk models from the measured ($S/N > 2$) or extrapolated
    ($S/N < 2$) stellar mass surface density profile. \label{rho}}
\end{center}
\end{figure*}

Both the surface brightness profiles and color profiles were derived 
from the binned ($6\times6$) star-subtracted images. Elliptical 
isophotes were fit using the IRAF task {\tt ellipse}, using
logarithmic spacing of ellipses. First, we determined the positions of
each isophote by running {\tt ellipse} on the summed $g+r$
image. Next, we ran {\tt ellipse} again on the individual images,
holding those positions fixed. Any nearby galaxies or imperfectly
subtracted stars were masked during this procedure.

To determine the sky
background, we measured the average flux in a circular annulus placed far from (but
centered on) each galaxy. The exact distance was chosen to maximize the separation from the
galaxy while avoiding any proximate galactic cirrus that might be present in the field
(we selected targets to be in regions of low cirrus; however due to Dragonfly's large FOV
this criterion does not exclude cirrus around the edges of the frame). To account for the
error in the sky level (a dominant source of uncertainty for the outer isophotes), we 
measured the average flux in multiple circular annuli over a range of distances and calculated the peak-to-peak scatter between those values. The error bars on the surface
brightness profiles therefore reflect this uncertainty as well as the error in the
photometric zeropoint determination and the variation in counts along each particular
isophote. Zeropoints were determined by calibrating to SDSS stars after correcting for
galactic foreground extinction.

Example $g$-band surface brightness profiles for NGC 1084 and NGC 4258
are shown in Figure \ref{egmug}. The surface brightness profiles reach
a signal-to-noise ratio of two at $30-32$ mag arcsec$^{-2}$ in $g-$band
(Table \ref{params} indicates the radius where this occurs for each galaxy). 
The full set of surface brightness profiles are shown in Figure \ref{mug}, and a
comparison between our profiles and previously published work is shown in the
Appendix. We note that one of the galaxies in our sample $-$ M101 $-$ has been
investigated in previous work with Dragonfly \citep{vanDokkum2014}; we reanalyzed
it here with the methods introduced in this paper for consistency. 

We converted the surface brightness profiles to
stellar mass surface density profiles (shown in Figure \ref{rho}) using the
relation between optical color and stellar mass-to-light ratios \citep{BelldeJong2001}
and assuming a Chabrier IMF \citep{Chabrier2003}. The error bars on
the surface density profiles include the intrinsic scatter associated
with this conversion ($0.1$ dex). The relation was determined from
galaxies in the SDSS DR7 \citep{Brinchmann2004} at low redshift
($0.045 < z < 0.055$) with similar stellar masses ($10 < {\rm
  log}_{10}M_{\rm stell} < 11$) and colors ($0.2 < g-r < 1.2$) as our
sample, and takes the following form \citep{vanDokkum2014}:  
\begin{equation}
{\rm log}\rho = -0.4(\mu_{g} - 29.23) + 1.49(g-r) + 4.58
\end{equation}

To account for Dragonfly's low spatial resolution, we replaced the
inner 2 kpc of each color profile with an extrapolation of an
exponential fit to the intermediate regions of the profile. We used the
observed colors beyond 2 kpc, until the signal-to-noise becomes too
low. The outermost regions of the color profile are a continuation of
the same exponential fit, with the exception of NGC 1084, for which we
assumed that the outer profile has the
same color as the innermost smooth component of the halo
(visible in the images).

\subsection{Stellar halos}
We define the stellar halo as the stellar mass ($M_{\rm stell}$) in
excess of a disk$+$bulge model, measured outside of 5 half-mass radii
($R_{h}$) (thus, the stellar halo fraction $f_{\rm halo}(>5R_{h}) =
M_{\rm stell}(>5R_{h})/M_{\rm stell}$). If a stellar halo exists in a
galaxy, it will only contribute significantly to the outermost regions
of the galaxy
\citep{Abadi2006,Johnston2008,Font2011,Cooper2013,Pillepich2015}, and
we therefore expect that a disk$+$bulge model derived from the high
surface brightness, inner regions (where we can visually verify that
the dominant contributions to the light profile are the disk and
bulge) will not be an adequate description for the lower surface
brightness, outer regions.  

\begin{figure*}[!t]
\begin{center}
\includegraphics[width=\textwidth]{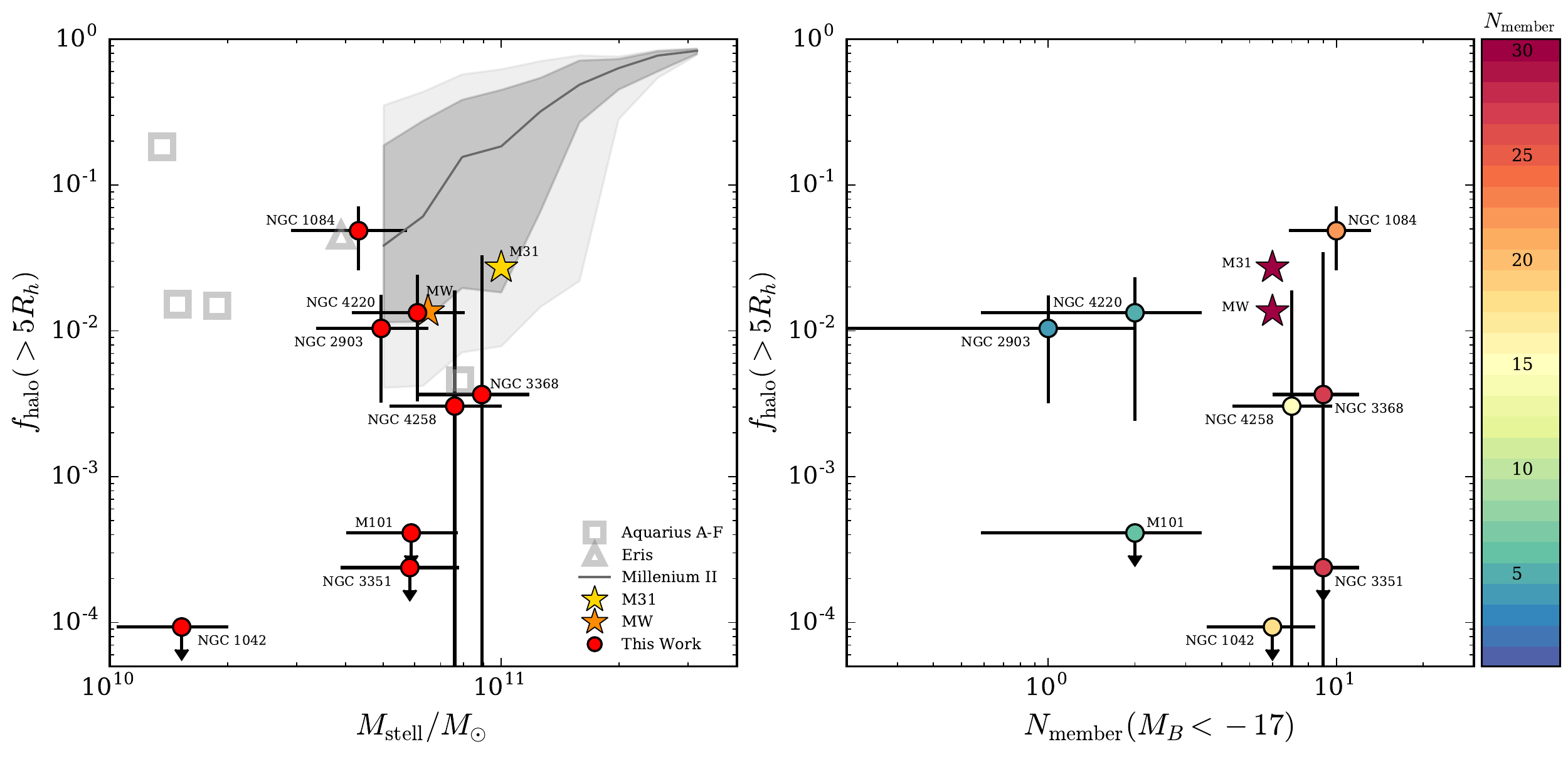}
\caption{
    \textit{Left}: The stellar halo mass fractions (and $1 \sigma$
    errors) for our sample, measured beyond $5R_{h}$ (red
    points). Values of $f_{\rm halo}$ for the Milky Way
    \citep{Carollo2010} and M31 \citep{Courteau2011} are shown for
    comparison (orange and gold stars, respectively), and have been
    scaled to the halo mass fraction outside of $5R_{h}$ assuming the
    structure of the halo of M31
    \citep{Irwin2005,Courteau2011}. Predictions of $f_{\rm halo}$, measured
    over $3 \leq r \leq 280$ kpc from the Aquarius simulations \citep{Cooper2010}; over
    $r \geq 20$ kpc from the Eris simulation \citep{Pillepich2015}; and over
    $r \geq 3$ kpc from the Millenium II simulation
    \citep[galaxies with B/T $< 0.9$ only;][]{Cooper2013} 
    are indicated by grey open squares, triangles, and
    shaded region, respectively.
    \textit{Right}: Environmental richness is parametrized by
    the number of group members \citep{MK2011} with $M_{B} < -17$. The
    color of each symbol corresponds to the \textit{total} number of
    known group members for that particular galaxy. The stellar halo
    mass fractions do not appear to be a function of
    environment.  \label{fhalo}}  
\end{center}
\end{figure*}

We therefore determine stellar halo masses
for each galaxy in the following way. First, we identified the
farthest extent of spiral arms in the unbinned image (in order to
utilize the higher spatial resolution; all subsequent measurements
were done on binned images), and fit an exponential disk model to the
stellar mass surface density profile over this region (Table \ref{params}). 
Next, we fit an
additional bulge model, keeping the sersic index $n$ as a free
parameter. To ensure the robustness of our fits and to obtain reliable
estimates of the uncertainties involved in this process, we used
\texttt{emcee} \citep{ForemanMackey2013}, an implementation of the
affine-invariant Monte Carlo Markov Chain (MCMC) ensemble sampler
\citep{GoodmanWeare2010}, to re-derive the maximum likelihood
model. This time we fit all five parameters  ($\Sigma_{0,\rm
  disk}$,$R_{s}$,$\Sigma_{0,\rm bulge}$,$R_{e}$,$n$)
simultaneously. We adopted a two-stage method to initialize the
\texttt{emcee} walkers. In a ``burn-in'' stage, we first initialized 
the walkers to small Gaussian distributions around our previous
best-fit parameters and ran the MCMC sampler. We then identified the
most likely solution from this run, and then ran the sampler again
after re-initializing the walkers to (tighter) Gaussians around the
new parameters. This allowed us to account for the possibility that
our original best-fit model differs significantly from the most likely
MCMC results, although we note that this step was unnecessary in most
cases. To quantify the $1\sigma$ uncertainties for the final disk$+$bulge
model, we sampled the posterior density functions (PDFs) $\sim 2000$
times and calculated the $16^{\rm th}$ and $84^{\rm th}$ percentiles
from the resulting set of models as a function of radius.

In addition to the bulge$+$disk model, we also fit a power law model
to the outskirts ($r > 7R_{s}$) of the stellar mass surface density
profiles (Figure \ref{rho}, applying the same methodology as before). These
fits were used to extrapolate the stellar mass surface density
profiles in our sample down to a uniform surface density of
$10^{4}M_{\odot}$ kpc$^{-2}$. The measured profiles were used where the $S/N$
exceeds 2, and this fit was used when the $S/N$ is lower than 2. We note that the
amount of mass in the extrapolated region is always small compared to the
total inferred halo mass; that is, the results are driven by the regions where
the $S/N$ exceeds 2 (except in the cases of NGC 1042 and NGC 1084; see below).
The final disk and bulge fits are shown alongside the stellar mass surface density profiles
for all galaxies in Figure \ref{rho}. Table \ref{params} contains the best-fit parameters
for the disk, bulge, and power law fits.

With these models in hand, we calculated the excess stellar mass as
the integral of the difference between the stellar mass surface
density profile (measured or extrapolated) and the disk$+$bulge model
outside of $5R_{h}$. The stellar halo mass fractions are shown in
Figure \ref{fhalo} and quoted in Table \ref{properties}. 

The average stellar halo and its error bar were computed using the
Kaplan-Meier product-limit estimator \citep{Feigelson1985}, a
nonparametric maximum-likelihood estimator commonly used in ``survival 
analysis'' (that is, the set of statistical techniques for analyzing
data that contains upper or lower limits). Using this method, we can
obtain an estimate of the true distribution sampled by our data points
as well as its mean and uncertainty.

To calculate the RMS scatter about the mean, we first calculate the
difference between the mean as determined by the Kaplan-Meier
estimator and the ``direct'' mean, that is, the mean that follows from
treating the halo fractions of the three galaxies with no detected
halo as measurements rather than upper limits. We then determined by
what factor the three upper limits need to be lowered so that the
direct mean equals the previously determined Kaplan-Meier estimator of
the mean. The RMS is calculated by applying this factor to the three
upper limits and then treating them in the same way as the other
measurements. This procedure ensures that the mean and the RMS are
obtained in a self-consistent way.

The average $f_{\rm halo}(>5R_{h})$ is 
$0.009 \pm 0.005$, and the RMS scatter is $1.01^{+0.09}_{-0.26}$ dex.
For galaxies that are consistent with having no mass in
the stellar halo (that is, $M_{\rm stell,halo} \sim 0$), we placed an
upper limit on the halo fraction by assuming the existence of a halo
with a constant surface density of $10^{4}M_{\odot}$ kpc$^{-2}$. The
excess stellar mass includes both smooth and structured halo
components, and early results from Dragonfly demonstrated that the
extensively studied halo of M31 would be readily detectable both in
the data and by impact on the radial profile out to $\sim 100$ kpc
\citep{vanDokkum2014}.  

We use halo fractions computed from surface density profiles
extrapolated down to $10^{4}M_{\odot}$ kpc$^{-2}$ in all cases except
for NGC 1084 and NGC 1042. A single power law is not an appropriate
model to fit to the outskirts of NGC 1084 since the surface density
profile steepens beyond $7R_{s}$, and therefore the extrapolation is a
clear overestimate (Figure \ref{rho}). NGC 1042 is a
non-detection, and its surface density profile reaches
$10^{4}M_{\odot}$ kpc$^{-2}$ at approximately $5R_{h}$, causing the
minimum extrapolated halo to be underestimated. We therefore consider
the directly measured halo fractions to be more reliable for these two
galaxies. If we were to use the extrapolated halo fractions for every
galaxy, we would obtain an average value of $0.009^{+0.005}_{-0.006}$
with an RMS scatter of $1.03^{+0.08}_{-0.24}$ dex.

Two of the galaxies in our sample (NGC 3368 and NGC 4258) have larger 
error bars that are consistent with a halo fraction of zero (Figure
\ref{fhalo}). The large uncertainties are driven by the fact that the 
visible excess above the disk$+$bulge model (Figure \ref{rho}) occurs
at substantially larger radii than $5R_{h}$. If we were to
measure the excess mass below a surface density threshold of
$10^{6}M_{\odot}$ kpc$^{-2}$ \citep[consistent with the profiles
obtained by e.g.,][]{Cooper2013} instead of outside of $5R_{h}$, the
halo fractions would become significant detections. Using this method
we would derive an average halo fraction of $0.006\pm 0.003$ with an
RMS scatter of $0.95^{+0.08}_{-0.22}$ dex, which, encouragingly, is
consistent with our original results. The decrease in the average halo
arises mainly from NGC 1084, for which the deviation from disk$+$bulge
occurs at surface densities higher than $10^{6}M_{\odot}$ kpc$^{-2}$.

\section{Discussion}
\subsection{A large variation in stellar halo mass fractions}

Confirming the visual impression of the galaxies in Figure \ref{allspirals}, we
find that the amount of mass in the stellar halo varies substantially between
galaxies in the sample, despite their small range in stellar mass
(Figure \ref{fhalo}). The stellar halo fractions have a peak-to-peak
span in $f_{\rm halo}(>5R_{h})$ of $>100$, ranging from $\leq 0.0001$
to $0.049 \pm 0.02$. The galaxy with the highest halo 
fraction is NGC 1084; this is also the only galaxy to feature a
prominent tidal stream in its halo \citep[consistent with previously
published imaging; ][]{MartinezDelgado2010}. At the other extreme,
three galaxies (NGC 1042, NGC 3351, and M101) are consistent  
with having {\em no} significant stellar halo, even at the detection
limits of Dragonfly. We note that our results for NGC 3351 are qualitatively consistent
with those of \cite{Watkins2014}, who report a lack of evidence for recent accretion based on
the exponential nature of the outer $B$-band surface brightness profile. 

The galaxies cover only a small range in stellar mass, but we do not find
a trend with mass within our sample (Figure \ref{fhalo}, left panel). This 
is consistent with halo light fractions based on
shallower observations of stacked individual galaxies with similar
stellar masses in Stripe82 \citep{Bakos2012}, although we measure a
larger scatter and lower mean for our sample. Correlations between
halo fraction and stellar mass are expected over broader ranges in
mass \citep{Pillepich2014}, and there is observational evidence that
stellar halos may become less common at lower stellar
masses than those considered here
(\citealt{Vlajic2011,BlandHawthorn2005,Streich2015}, but see also
\citealt{Belokurov2016}). Moreover, the stellar halo fractions do not
show any environmental dependence (right panel of Figure
\ref{fhalo}). Environment for each galaxy is parametrized by the number of
bright ($M_{B} < -17$) associated group members, as reported in the MK
group catalog \citep{MK2011}. Table \ref{properties} contains the number of group
members for each galaxy. This lack of a trend between stellar halo
fraction and environmental richness has also been seen in simulations
\citep{Lackner2012}.     

\subsection{Inclination effects}
Edge-on galaxies are generally considered ideal laboratories for
stellar halo studies, as contamination from stars in the main disk and
bulge is expected to fall off rapidly along the minor axis. As we did
not select galaxies based on any intrinsic property other than
luminosity, in our small sample we do not have any entirely edge-on
galaxies. We extracted measuresments of inclination angles from the
HyperLEDA database \citep{LEDA}. 

The average inclination angle for
galaxies in our full sample is $56.8^{\circ}$, and galaxies with and
without detected halos have average inclinations of $65.2$ and $42.9$,
respectively. This is not necessarily surprising, since detecting 
faint halos is substantially more difficult to do in the presence of a
relatively bright disk, and it is possible that the upper limits we
derived for the three galaxies with non-detections are artificially
lower as a consequence of this effect. However, we emphasize that the
maximum likelihood disk$+$bulge model was determined in regions
significantly interior to where the stellar halo mass was measured and
we are therefore not relying on any ability to perform disk-halo
decompositions at low surface density (and in fact, with the exception of
M101, the galaxies for which no halo has been detected show a
noticeable \textit{deficit} of stellar mass surface density relative
to the disk$+$bulge model). Furthermore, while on average the galaxies
with detected halos are closer to being edge-on, the individual
measurements range from $49.9^{\circ}$ to $\sim90^{\circ}$, showing
clear overlap with the sample with non-detections ($16.1^{\circ} \leq
i \leq 58^{\circ}$) and demonstrating our ability to detect an excess
of stellar mass even in only moderately inclined galaxies.

\subsection{Outer disk morphology}
Our method for measuring stellar halos relies on the assumption that the stellar mass surface density of the disk of the galaxy follows an exponential profile. This is a simplification, of course, as disks are known to frequently display more complex morphologies. 

The surface brightness profiles of disk galaxies are only rarely a pure exponential (Type I) profile $-$ approximately $60\%$ of disks show a break in the radial profile at approximately $2.5R_{\rm disk}$, followed by a second, steeper exponential (Type II), and the remaining $30\%$ (Type III) have a break followed by a shallower exponential \citep{vanderKruit1979,PohlenTrujillo2006,Laine2014}. Furthermore, \cite{MartinNavarro2012} showed that truncations are also common in disks \citep[at $3-5 R_{\rm disk}$;][]{vanderKruitSearle1982,BarteldreesDettmar1994,Pohlen2000}, although they occur at greater radial distances (lower surface brightness) and are thus more difficult to observe in face-on disks than in edge-on disks. 

Stellar mass surface density profiles are less complicated. \cite{Bakos2008} demonstrated that both Type I and Type II disks have exponential stellar mass surface density profiles devoid of any significant break \citep[a consequence of the characteristic ``U'' -shaped optical color profiles for Type II galaxies; see also][]{Zheng2015,RuizLara2016}. Truncations are still present, however, and the surface densities at which they occur are comparable to the surface densities at which stellar halos begin to dominate the total profile \citep{Bakos2012,MartinNavarro2014}.

%XXXXX reminder that this is still about 27 mag per arcsec2 and well above our image depth... XXXXX

Of the three galaxies in our sample that do not appear to have stellar halos, one has a stellar mass surface density that is a pure exponential (M101), and the others (NGC 1042 and NGC 3351) show truncations at $4.8 R_{\rm disk}$ and $4.5 R_{\rm disk}$, respectively. It is possible that each of these galaxies harbors a faint stellar halo that is either exactly balancing the drop in surface density due to a truncation (M101) or making the observed truncations shallower. 

To investigate what the maximum halo contribution could be, we compute the stellar halo masses
once more, assuming that the halos contain \textit{all} the stellar mass beyond the
last visible spiral arms. By definition, every galaxy in our sample now has a 
significant stellar halo, and the average halo fraction for the sample is $0.04 \pm 0.03$.
Importantly, the RMS scatter in halo fractions using this method is still very high, at
$0.84^{+0.15}_{-0.36}$ dex. We are therefore confident that our primary finding of a
substantial scatter between the stellar halos of massive spiral galaxies is robust to our
assumptions of an exponential disk.

\subsection{Thick Disks?}
Further complications arise on the observational side, due to the
possible contribution from thick disks, an additional, relatively low
surface density component that ubiquitously exists in disk galaxies
\citep{Dalcanton2002}. It is possible, considering the disky
isophotes of the low surface brightness regions of certain galaxies in
our sample (Figure \ref{allspirals}), that we have detected some thick
disks and misinterpreted them as stellar halos. However, the mass in thick
disks detected in external galaxies has been typically reported to be
roughly $10$\% of the mass of the thin disk \citep{Yoachim2006}, and
more recent estimates have increased this to be nearly comparable to
the mass of the thin disk in some cases \citep{Comeron2011}, whereas the
excess mass - to - disk mass ratios in our sample are are all $\lesssim
5$\%, with the majority $\lesssim 1$\%. 

If the excess mass we measure
\textit{were} part of the thick disk, this would complicate inferences
about the accretion histories of these galaxies. The formation
mechansim of thick disks \citep[or the stratified populations which mimic them, as suggested by][]{BovyRixHogg2012} is still debated, but competing theories
involve internal disk heating from interactions with spiral arms or
star clusters \citep{Loebman2011}, formation as a thick disk at high
redshift from material acquired during gas-rich mergers
\citep{Brook2004}, buildup from shredded satellites accreted at low
impact angle \citep{Penarrubia2006,Read2008}, disk heating by massive
accretion events \citep{Abadi2003,Read2008}, or, more likely, a
combination of the above. Our data do not allow us to distinguish
between these possibilities, but it is reasonable to assume that if we
have observed thick disks, at least some of the stellar mass was
accreted or indicative of a massive past accretion event. 

More to the point, if part of the excess light is due to a thick disk component
rather than a stellar halo, the mean halo mass fraction of the sample would be
even lower. This in turn would strengthen our conclusion that M31 in particular
has had an unusually active recent accretion history relative to other massive
spiral galaxies.

\subsection{Stellar halo fractions in context}
Under the assumption that stellar halos are built primarily through
the aggregation of individual accretion events over time, the large
variation observed in halo mass fractions is reflective of the underlying
variation in accretion histories of spiral
galaxies. The accretion histories of the
Milky Way and M31 are already suspected to have been significantly
different, based on (among other metrics) the amount and global
structure of light in their stellar
halos \citep{Carollo2010,Courteau2011,Deason2013}. Similarly, from the
halo fraction (and visible presence of a tidal stream) we can infer 
that NGC 1084 has had a relatively active accretion history that
continues to influence its growth. 

It is possible that, given the low spatial resolution of Dragonfly
($2.85$ arcsec pixel$^{-1}$), the three galaxies that have no detected
halos harbor faint, thin streams that are undetectable in our data, or
that the stellar halos are diffuse enough to never surpass the surface
density of the disk even at large radii. Still, our results place
strong constraints on the possible nature of the accretion histories
of these three galaxies. Specifically, we can rule out massive
($>10^{7}M_{\rm  stell}$) accretion events over the past $\sim 4.5$
Gyr that would have resulted in a significant and irregularly
structured halo, as well as a substantial and continuous infall of low
mass systems, as this would produce a shallow power law profile
similar to that of M31 \citep{Deason2013}. 

The growth of stellar halos is known to be a stochastic process
\citep{Amorisco2015}, and simulations predict an RMS scatter in halo
fractions of $0.5-0.6$ dex \citep[e.g.,][see Figure \ref{fhalo}]{Cooper2010,Cooper2013}. The stellar halos of the eight spirals presented here have an RMS
scatter of $1.01^{+0.09}_{-0.26}$ dex, approximately $2-3\times$ higher than theoretical expectations. Additionally, the average halo fraction for
our sample is $0.009 \pm 0.005$, slightly lower than predicted
values.  

Such direct comparisons of average halo fractions between simulations
and observations can be misleading, however, as the details of the simulation can have a substantial effect on the properties of the stellar halo. N-body simulations, for example, are significantly faster than hydrodynamic simulations and therefore have the advantage of being able to easily build up large samples of galaxies and stellar halos (a requirement for studying the intrinsic scatter in accretion histories); the trade-off is that any non-gravitational effects are neglected, and assigning stellar mass to DM particles is non-trivial and can lead to systematic changes in the concentration, morphology, extent, and amount of structure in stellar halos by factors of $2-7$ \citep{Bailin2014}.

Moreover, the stellar halo mass fraction and the accreted stellar mass fraction are
not identical quantities. Two of the three simulations that we compare our results to
in Figure \ref{fhalo} \citep{Cooper2010,Cooper2013} produce accretion-only stellar halos, but evidence (from observations and simulations alike) suggests that stars can come to reside in the stellar halo by means of in-situ star formation \citep{Zolotov2009,Sheffield2012}, stellar migration \citep{RadburnSmith2012}, or ejection \citep{Purcell2010,Dorman2013} \citep[although beyond $20$ kpc accreted stars are expected to contribute $\sim97$\% of the total stellar mass, according to][]{Pillepich2015}. 

The majority of the mass in stellar halos is expected to be contributed by only a few relatively massive accretion events, which ultimately merge with the central galaxy and deposit most of their stars onto the inner regions \citep{Deason2013,Pillepich2015}. Inconsistencies in stellar halo definitions can therefore introduce uncertainties into any comparison between observations and simulations or between any two simulations. Of the three simulations that we compare to in Figure \ref{fhalo}, \cite{Cooper2010} define the stellar halo as all stellar mass between $3$ and $280$ kpc; \cite{Cooper2013} measure the stellar mass beyond $3$ kpc; and \cite{Pillepich2014} measure the stellar mass beyond $20$ kpc. Our method of calculating the stellar mass in excess of a disk$+$bulge beyond $5R_{h}$ results in a different physical radius for each galaxy in our sample, with values ranging from $11$ to $32$ kpc (Table \ref{properties}).

By design, our calculation of $f_{\rm halo}(>5R_{h})$ does not rely on any assumptions about the global shape of the stellar halo, due to a lack of constraining power in the innermost regions. In the future, a more detailed comparison with simulations might therefore be made by choosing a set of structural parameters, and applying a scaling factor $\kappa = f_{\rm halo} / f_{\rm  halo}(>5R_{h})$; doing so would systematically change the halo fractions, but it is unlikely that the RMS scatter would decrease after such a rescaling.

\section{Conclusions}
We have presented measurements of the stellar halo mass fraction, defined as the stellar mass in excess of a disk$+$bulge model outside of $5R_{h}$, for eight nearby spiral galaxies. Considering the relatively narrow range in stellar mass ($2-8\times 10^{10}M_{\odot}$), we find a remarkabley wide range in stellar halo mass fractions. One of the galaxies in our sample, NGC 1084, has a stellar halo mass fraction of $0.049 \pm 0.02$, while three others (NGC 1042, NGC 3351, and M101) have stellar halos that are undetected in our data. We measure an RMS scatter of $1.01^{+0.09}_{-0.26}$ dex, and a peak-to-peak span of a factor of $>100$. Placing tighter constraints on the three galaxies that appear to be without stellar halos requires even deeper imaging than what we are presenting here. This will be extremely challenging, but it may be possible with Dragonfly. 

The variation in the masses of stellar halos that we find (and implied levels of stochasticity in the accretion histories of these galaxies) is qualitatively consistent with variations in the structure and stellar populations of nearby stellar halos observed in both integrated light and star counts studies \citep{Mouhcine2007,Tanaka2011,Barker2012,Monachesi2015}, although these efforts typically suffer from PSF effects \citep{deJong2008,Slater2009,Sandin2014,Duc2015} and sparse area coverage, respectively. 

Looking forward, a more comprehensive understanding of the buildup of stellar halos will require not only improved techniques for robust comparisons between observations and simulations, but also a much larger sample of observed stellar halos in order to capture the full extent of the scatter at fixed stellar mass. The advantage of using Dragonfly to study the global properties of stellar halos lies in the combination of its sensitivity, large field of view and relative lack of susceptibility to scattered starlight. Deep integrated light surveys are complementary to resolved star count studies that supply information on the extent and stellar populations of stellar halos in pencil-beam surveys \citep[e.g., GHOSTS;][]{Radburnsmith2011}. In the future, wide field space-based imaging (with telescopes such as WFIRST) coupled with kinematic data will enable us to characterize the structure and content of a larger sample of stellar halos in the same detail as is possible in the Local Group today.

\acknowledgments
We thank the anonymous referee for an insightful and thorough report that improved the paper.
Support from NSERC, NSF grant AST-1312376 and from the Dunlap Institute (funded by the
David Dunlap Family) is gratefully acknowledged. All
authors thank the staff at New Mexico Skies Observatory for their
support and assistance; Andrew Cooper for providing the data necessary
to construct the models in Figure \ref{fhalo}; and the PAndAS team for providing
their M31 star count data. AM thanks the BS group for useful discussions and insights.

\newpage

\appendix

\section{Photometric Tests}
All of the galaxies in our sample are well known, and well studied. We therefore carried out
careful comparisons with previously published surface brightness profiles to ensure that our
profiles are consistent with existing photometry.

We went about this in two ways $-$ first, we compared directly to profiles found in the literature. As discussed in the main text, M101 has been studied previously by e.g. \cite{Mihos2013} and \cite{vanDokkum2014}. \cite{PohlenTrujillo2006} published $g$-band surface brightness profiles of NGC 1042 and NGC 1084; \cite{Watkins2014} derived $B$-band surface brightness profiles for NGC 3351 and NGC 3368; and a $B$-band profile of NGC 4258 is presented in \cite{Watkins2016}. 

\begin{figure*}[!t]
\begin{center}
\includegraphics[width=\textwidth]{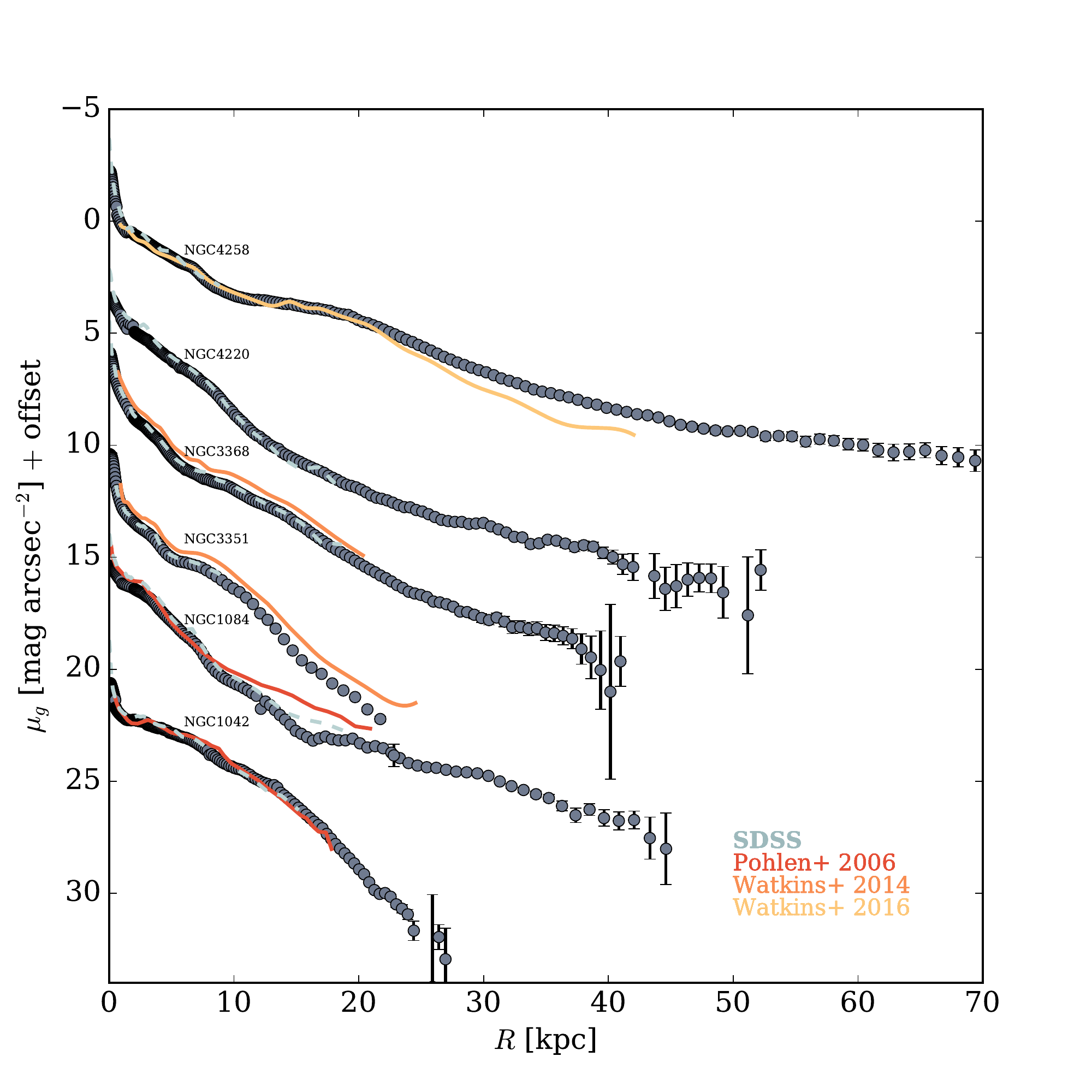}
\caption{A comparison between Dragonfly $g$-band surface brightness profiles (grey points)
and profiles obtained from the literature. Arbitrary offsets have been applied to each
profile to avoid confusion. Additional comparisons to SDSS data are shown as blue dashed
lines. We find good agreement with our photometry in both cases.
 \label{compfig}}  
\end{center}
\end{figure*}

Figure \ref{compfig} shows our surface brightness profiles with these literature profiles superimposed (we applied arbitrary offsets to each profile to avoid confusion between
different galaxies). Where existing profiles are measured in $B$-band, we use the available 
$B-V$ profiles to convert to $g$-band, applying conversions from \cite{Blanton2007}. The
agreement between Dragonfly and literature profiles is generally good, with the exception of
NGC 3351 and 3368, for which the profiles presented in \cite{Watkins2014} appear to have a
systematic offset of 0.5 magnitudes. We note that the apparent disagreement between our
results and those of \cite{Watkins2016} for NGC 4258 at large radii is due to differences in
averaging; \cite{Watkins2016} measure the median flux in elliptical annuli, whereas we
measure the mean.

As a second, independent approach, we also derived surface brightness profiles for the
(relatively) high surface brightness regions of the galaxy using SDSS images; these are
indicated in Figure \ref{compfig} as blue lines. We find very good agreement here
between the SDSS and Dragonfly profiles. In particular, we note that the surface brightness
profiles of NGC 3351 and NGC 3368, which are somewhat in tension with the converted $B$-band
profiles of \cite{Watkins2014}, are consistent with the SDSS $g$-band surface brightness
profiles.

%* M101 - van Dokkum 2014 (g), Mihos 2013 (B)
%* NGC 1042 - Pohlen & Trujillo 2006 (sdss g,r)
%* NGC 1084 - Pohlen & Trujillo 2006 (sdss g,r)
%* NGC 3368 / M96 - Watkins 2014 (B) 
%* NGC 3351 / M95 - Watkins 2014 (B) 
%* NGC 4258 / M106 - Watkins 2016 (B)
%* Conversions from g to B done via Blanton 2007
%* g-r vs B-V done via Fukugita 1996further details? maybe? XXXX

\vspace{6em}

%\bibliography{spirals}

%%%% CAPTIONS FOR TABLES ARE EDITED IN F_HALO.PY !!!!!!!!!!!

\begin{table*}  
\centering  
\rotatebox{90} {  
\hskip-1.0cm 
\begin{tabular}{ccccccccc}  
\hline  
${\rm Galaxy \,\,Name}$ &  $\alpha$ & $\delta$ & $M_{B}\,\,[\rm mag]$ & $D \,\,[\rm Mpc]$ & $N_{\rm member}$ & $M_{\rm stell}\,\,[M_{\odot}]$ & $5R_{h}$ [kpc] & $f_{\rm halo}(>5R_{h})$ \\ 
\hline 
${\rm NGC\,1042}$ & $2:40:23.97$ & $-8:26:01.13$ & $-20.27$ & $17.3$ & $6$ & $1.53\pm0.48\times 10^{10}$ & 26.6 & $0.0001\pm0.0001$ \\ 
${\rm NGC\,1084}$ & $2:45:59.92$ & $-7:34:43.10$ & $-20.41$ & $17.3$ & $10$ & $4.32\pm1.41\times 10^{10}$ & 17.0 & $0.049\pm0.023$ \\ 
${\rm NGC\,2903}$ & $9:32:10.11$ & $+21:30:02.99$ & $-20.3$ & $8.5$ & $1$ & $4.94\pm1.57\times 10^{10}$ & 19.5 & $0.0104\pm0.0072$ \\ 
${\rm NGC\,3351}$ & $10:43:57.73$ & $+11:42:13.00$ & $-20.36$ & $10.0$ & $9$ & $5.84\pm1.95\times 10^{10}$ & 10.4 & $0.0002\pm0.0225$ \\ 
${\rm NGC\,3368}$ & $10:46:45.74$ & $+11:49:11.78$ & $-20.03$ & $7.24$ & $9$ & $8.91\pm2.87\times 10^{10}$ & 12.8 & $0.0037\pm0.0292$ \\ 
${\rm NGC\,4220}$ & $12:16:11.73$ & $+47:53:00.07$ & $-19.31$ & $17.1$ & $2$ & $6.11\pm1.96\times 10^{10}$ & 13.7 & $0.0133\pm0.0109$ \\ 
${\rm NGC\,4258}$ & $12:18:57.62$ & $+47:18:13.39$ & $-20.2$ & $7.61$ & $7$ & $7.61\pm2.42\times 10^{10}$ & 20.8 & $0.003\pm0.0159$ \\ 
${\rm M101}$ & $14:03:12.58$ & $+54:20:55.50$ & $-20.2$ & $7.0$ & $2$ & $5.89\pm1.87\times 10^{10}$ & 32.7 & $0.0004\pm0.0008$ \\ 
\hline 
\end{tabular} 
} 
\vspace{40pt}  
\caption{Properties of the sample. The stellar halo fractions for M101, NGC 1042 and NGC 3351 are upper limits. Magnitudes and distances were obtained from the Extragalactic Distance Database \cite{Tully2009}; group members were obtained from the MK Groups catalog \cite{MK2011}. \label{properties}}   
\end{table*}

\begin{table*}  
\centering  
\rotatebox{90} {  
\hskip-1.0cm 
\begin{tabular}{cccccccccc}  
\hline  
${\rm Galaxy \,\,Name}$ &  $R_{\rm arm}$ [kpc] & $R_{2}$ [kpc] & $A_{d}$ [$M_{\odot}$ kpc$^{-2}$] & $R_{s}$ [kpc] & $A_{b}$ [$M_{\odot}$ kpc$^{-2}$] & $R_{e}$ [kpc] & $n$ & $A$ & $B$ \\ 
\hline 
${\rm NGC\,1042}$ & $15.0$ & $24.9$ & $3.9^{+0.2}_{-0.2} \times 10^{8}$ & $3.0^{+0.1}_{-0.1}$ & $1.3^{+0.8}_{-0.3} \times 10^{9}$ & $1.1^{+0.2}_{-0.1}$ & $0.9^{+0.4}_{-0.2}$ & $-9.5^{+1.9}_{-1.3}$ & $17.4^{+1.7}_{-2.6}$ \\ 
${\rm NGC\,1084}$ & $6.0$ & $43.3$ & $2.3^{+0.4}_{-0.4} \times 10^{9}$ & $2.0^{+0.2}_{-0.2}$ & $3.2^{+1.4}_{-1.3} \times 10^{9}$ & $2.3^{+1.1}_{-0.9}$ & $1.3^{+0.5}_{-0.5}$ & $-3.3^{+0.2}_{-0.2}$ & $10.4^{+0.3}_{-0.3}$ \\ 
${\rm NGC\,2903}$ & $20.0$ & $40.6$ & $1.8^{+0.2}_{-0.2} \times 10^{9}$ & $2.9^{+0.1}_{-0.1}$ & $1.0^{+0.1}_{-0.1} \times 10^{10}$ & $1.5^{+0.4}_{-0.3}$ & $0.8^{+0.2}_{-0.2}$ & $-6.3^{+0.5}_{-0.5}$ & $14.7^{+0.7}_{-0.7}$ \\ 
${\rm NGC\,3351}$ & $12.0$ & $22.8$ & $7.5^{+2.9}_{-2.9} \times 10^{8}$ & $2.9^{+0.5}_{-0.3}$ & $6.8^{+4.1}_{-2.1} \times 10^{10}$ & $1.7^{+0.2}_{-0.2}$ & $1.7^{+0.4}_{-0.3}$ & $-8.7^{+2.7}_{-1.8}$ & $16.6^{+2.4}_{-3.7}$ \\ 
${\rm NGC\,3368}$ & $12.0$ & $37.1$ & $4.5^{+3.5}_{-1.8} \times 10^{8}$ & $4.3^{+1.0}_{-0.8}$ & $8.1^{+0.9}_{-1.4} \times 10^{10}$ & $2.7^{+0.2}_{-0.2}$ & $1.9^{+0.1}_{-0.2}$ & $-4.6^{+1.0}_{-1.0}$ & $12.3^{+1.6}_{-1.5}$ \\ 
${\rm NGC\,4220}$ & $4.5$ & $41.2$ & $3.3^{+0.5}_{-1.1} \times 10^{9}$ & $2.3^{+0.3}_{-0.1}$ & $5.1^{+1.3}_{-1.2} \times 10^{9}$ & $2.6^{+2.3}_{-1.0}$ & $0.9^{+0.4}_{-0.3}$ & $-4.0^{+0.2}_{-0.2}$ & $11.3^{+0.2}_{-0.2}$ \\ 
${\rm NGC\,4258}$ & $28.0$ & $69.4$ & $6.0^{+0.5}_{-0.5} \times 10^{8}$ & $5.3^{+0.1}_{-0.1}$ & $2.5^{+0.0}_{-0.1} \times 10^{10}$ & $3.9^{+0.0}_{-0.1}$ & $1.7^{+0.1}_{-0.1}$ & $-3.9^{+0.4}_{-0.4}$ & $11.8^{+0.6}_{-0.6}$ \\ 
${\rm M101}$ & $40.0$ & $41.9$ & $5.2^{+0.3}_{-0.4} \times 10^{8}$ & $4.1^{+0.1}_{-0.1}$ & $9.0^{+4.9}_{-2.8} \times 10^{9}$ & $1.5^{+0.3}_{-0.3}$ & $1.9^{+0.5}_{-0.4}$ & $-9.5^{+0.5}_{-0.2}$ & $19.5^{+0.4}_{-0.8}$ \\ 
\hline 
\end{tabular} 
} 
\vspace{40pt}  
\caption{Parameters from the best fit disk+bulge model and power law extrapolation. The projected radii at which the last visible spiral arms are located and the signal to noise ratio drops below 2 are given by $R_{\rm arm}$ and $R_{2}$, respectively.  \label{params}}  
\end{table*}

\end{document}